\documentclass{aastex}
\usepackage{emulateapj5}

\usepackage{amsbsy}

\usepackage{psfig}

\newcommand{\mbf}[1]{\mbox{\boldmath $#1$}}

\newcommand{\boost}{\ensuremath{{\bf B}\mbf{_{\hat m}}(\beta)}}
\newcommand{\rotat}{\ensuremath{{\bf R}\mbf{_{\hat n}}(\phi)}}
\newcommand{\rotEn}{\ensuremath{{\bf R}\mbf{_{\hat n}}^{(3)}(2\phi)}}

\newcommand{\rotq}{\ensuremath{{\bf R}\mbf{_{\hat q}}(\Phi_I)}}
\newcommand{\rotv}{\ensuremath{{\bf R}\mbf{_{\hat v}}(\Delta\Psi)}}
\newcommand{\rotEq}{\ensuremath{{\bf R}\mbf{_{\hat q}}^{(3)}(2\Phi_I)}}
\newcommand{\rotEv}{\ensuremath{{\bf R}\mbf{_{\hat v}}^{(3)}(2\Delta\Psi)}}

\newcommand{\psr}{PSR\,J0437$-$4715}

\shorttitle   {Phase-Coherent Matrix Convolution}
\shortauthors {W. van Straten}

\begin{document}

\title{Radio Astronomical Polarimetry and Phase-Coherent Matrix Convolution}

\author{W. van Straten}

\affil{Centre for Astrophysics and Supercomputing,
	Swinburne University of Technology, \\
	Hawthorn, VIC 3122, Australia}

\email{wvanstra@pulsar.physics.swin.edu.au}

\begin{abstract}

A new phase-coherent technique for the calibration of polarimetric
data is presented.  Similar to the one-dimensional form of
convolution, data are multiplied by the response function in the
frequency domain.  Therefore, the system response may be corrected
with arbitrarily high spectral resolution, effectively treating the
problem of bandwidth depolarization.  As well, the original temporal
resolution of the data is retained.  The method is therefore
particularly useful in the study of radio pulsars, where high time
resolution and polarization purity are essential requirements of
high-precision timing.  As a demonstration of the technique, it is
applied to full-polarization baseband recordings of the nearby
millisecond pulsar, \psr.

\end{abstract}


\section {Introduction}

In radio polarimetry, two orthogonal senses of polarization are
received and propagated through separate signal paths.  Each signal
therefore experiences a different series of amplification,
attenuation, mixing, and filtering before sampling or detection is
performed.  Whereas efforts are made to match the components of the
observatory equipment, each will realistically have a unique frequency
response to the input signal.  Even a simple mismatch in signal path
length will result in a relative phase difference between the two
polarizations that varies linearly with frequency.

In fact, any physically realizable system will transform the radiation
in a frequency-dependent manner.  Where variations across the smallest
bandwidth available may be considered negligible, post-detection
calibration and correction techniques may be used to invert the
transformation and recover the original polarimetric state.  However,
the transformation may vary significantly across the band, causing the
polarization vector to combine destructively when integrated in
frequency.  This phenomenon is known as ``bandwidth depolarization''
of the signal, and results in irreversible decimation of the degree of
polarization.

It is therefore desirable to perform polarimetric corrections at
sufficiently high spectral resolution, which is available only at the
cost of temporal resolution unless phase coherence is maintained.  
Conventional post-detection correction techniques therefore prove
insufficient when high time resolution is also a necessity.  For
example, certain pulsar experiments require high time resolution in
order to resolve key features in the average pulse profile.  As well,
it has been shown that insufficient time resolution can also lead to
depolarization \citep{gxv+99}.  These considerations motivate the
development of a method for phase-coherent polarimetric
transformation.  Further impetus is provided by the growing number of
baseband recording systems at observatories around the world, the
enhanced flexibility made available through use of off-line data
reduction software, and the increasing computational power of
affordable facilities.

A method has previously been presented for the phase-coherent
correction of interstellar dispersion smearing.  Convolution is
performed by multiplying the spectrum of baseband data with the
inverted frequency response of the interstellar medium (ISM), as
modeled by the cold plasma dispersion relation \citep{hr75}.  The
current development extends the concept of convolution to
multiplication of the vector spectrum by the inverse of the frequency
response matrix.  The term ``matrix convolution'' is used to
distinguish this operation from two-dimensional convolution, such as
would be performed on an image.  Whereas the frequency response matrix
may be composed of various factors, such as those arising from the ISM
and ionosphere, this paper will be restricted to considering only the
instrumental response.

Following a brief review of matrix convolution and its relationship to
the transformation properties of radiation, a technique is proposed by
which the instrumental frequency response matrix may be determined and
used to calibrate phase-coherent baseband data.  As a preliminary
illustration, the method is applied to full-polarization observations
of the millisecond pulsar, \psr, from two different receiver systems.
The results agree well with previously published polarimetry for this
pulsar \citep{nms+97}.

\section{Matrix Convolution}
\label{sec:convolution}

Consider a linear system with impulse response, $j(t)$.  Presented
with an input signal, $e(t)$, the output of this system is given by
the convolution, $e^\prime(t)=j(t)*e(t)$.  In the two-dimensional
case, each output signal is given by a linear combination of the input
signals,
\begin{eqnarray}
e_1^\prime(t)=j_{11}(t)*e_1(t) + j_{12}(t)*e_2(t), \\
e_2^\prime(t)=j_{21}(t)*e_1(t) + j_{22}(t)*e_2(t).
\end{eqnarray}
Now let $e_1(t)$ and $e_2(t)$ be the complex-valued analytic signals
associated with two real time series, providing the instantaneous
amplitudes and phases of two orthogonal senses of polarization.  By
defining the analytic vector, $\mbf{e}(t)$, with elements $e_1(t)$ and
$e_2(t)$, and the $2\times2$ impulse response matrix, ${\bf{j}}(t)$,
with elements $j_{mn}(t)$, we may express the propagation of a
transverse electromagnetic wave by the matrix equation,
\begin{equation}
{\mbf{e^\prime}}(t)={\bf{j}}(t)*\mbf{e}(t).
\label{eqn:convolution_t}
\end{equation}
By the convolution theorem, equation~(\ref{eqn:convolution_t}) is
equivalent to
\begin{equation}
\mbf{E^\prime}(\nu)={\bf J}(\nu)\mbf{E}(\nu),
\label{eqn:convolution_nu}
\end{equation}
where $\bf{J}(\nu)$ is the frequency response matrix with elements
$J_{mn}(\nu)$, and $\mbf{E}(\nu)$ is the vector spectrum.  In the case
of monochromatic light, or under the assumption that $\bf{J}(\nu)$ is
constant over all frequencies, matrix convolution reduces to simple
matrix multiplication in the time domain, as traditionally represented
using the Jones matrix.  However, because these conditions are not
physically realizable, the Jones matrix finds its most meaningful
interpretation in the frequency domain.

The average auto- and cross-power spectra are neatly summarized
by the average power spectrum matrix, defined by the vector direct
product, $\bar{\bf{P}}(\nu)
=\langle\mbf{E}(\nu)\boldsymbol{\otimes}\mbf{E}^\dagger(\nu)\rangle$,
where $\mbf{E}^\dagger$ is the Hermitian transpose of $\mbf{E}$ and
the angular brackets denote time averaging.  More explicitly:
\begin{equation}
\bar{\bf{P}}(\nu)=
  \left( 
    \begin{array}{cc}
      \langle E_1(\nu)E_1^*(\nu)\rangle & \langle E_1(\nu)E_2^*(\nu)\rangle \\
      \langle E_2(\nu)E_1^*(\nu)\rangle & \langle E_2(\nu)E_2^*(\nu)\rangle
    \end{array}
  \right).
\label{eqn:avg_power_spectrum_matrix}
\end{equation}
Each component of the average power spectrum matrix,
$\bar{P}_{mn}(\nu)$, is the Fourier transform pair of the average
correlation function, $\bar{\rho}_{mn}(\tau)$ \citep{pap65}.
Therefore, $\bar{\bf{P}}(\nu)$ may be related to the commonly used
coherency matrix,
\begin{equation}
\boldsymbol{\rho}=
	\langle\mbf{E}(t)\boldsymbol{\otimes}\mbf{E}^\dagger(t)\rangle
=\bar{\boldsymbol{\rho}}(0)
={1\over2\pi}\int_{\nu_0-\Delta\nu}^{\nu_0+\Delta\nu}{\bar{\bf{P}}(\nu)}d\nu,
\label{eqn:coherency_matrix}
\end{equation}
where $\nu_0$ is the centre frequency and $2\Delta\nu$ is the bandwidth
of the observation.  The average power spectrum matrix may therefore
be interpreted as the coherency spectral density matrix and, in the limit
$\Delta\nu\rightarrow0$, $\boldsymbol{\rho}=\bar{\bf{P}}(\nu_0)/2\pi$.

Using equations~(\ref{eqn:convolution_nu})
and~(\ref{eqn:avg_power_spectrum_matrix}) it is easily shown that a
two-dimensional linear system transforms the average power spectrum as
\begin{equation}
{\bar{\bf{P}}^\prime}(\nu)={\bf{J}}(\nu)\bar{\bf{P}}(\nu){\bf{J}}^\dagger(\nu).
\label{eqn:congruence_transformation}
\end{equation}
This matrix equation is called the congruence transformation, and
forms the basis on which the frequency response of a system will be
related to the input (source) and output (measured) coherency
spectrum.  For brevity in the remainder of this paper, all symbolic
values are assumed to be a function of frequency, $\nu$, unless
explicitly stated otherwise.

\section{Isomorphic Representations}
\label{sec:isomorphisms}

The subject of radio astronomical polarimetry received its most
rigorous and elegant treatment with the two separate developments of
\citet{bri00} and \citet{ham00}.  Whereas Britton illuminates the
isomorphism between the transformation properties of radiation and
those of the Lorentz group, Hamaker illustrates and utilizes the
similarity with the multiplicative quaternion group.  The salient
features of both formalisms provide a sound foundation for the current
development, and merit a brief review.

Any complex 2$\times$2 matrix, {\bf J}, may be expressed as
\begin{equation}
{\bf J} = J_0{\bf I} + \mbf{J\cdot\sigma},
\label{eqn:quaternion}
\end{equation}
where $J_0$ and $\mbf{J}=(J_1,J_2,J_3)$ are complex, {\bf{I}} is the
$2\times2$ identity matrix, and $\boldsymbol{\sigma}$ is a 3-vector
whose components are the Pauli spin matrices.  The 4-vector,
[$J_0,\mbf{J}$], may be treated using the same algebraic rules as
those used for quaternions. In the remainder of this paper, the
equivalent quaternion and 2$\times$2 matrix forms (as related by
eq.~[\ref{eqn:quaternion}]) will be interchanged freely.

If {\bf J} is Hermitian, then the components of [$J_0,\mbf{J}$] are
real.  The average power spectrum and coherency matrices are Hermitian
and, when decomposed in this manner, 2[$S_0,\mbf{S}$] may be
interpreted as the mean Stokes parameters, where $2S_0$ is the total
intensity and $2\mbf{S}$ is the polarization vector.  By writing
${\bar{\bf{P}}}=[\bar{S}_0,\bar{\mbf{S}}]$, it is more easily seen
that the integration of equation~(\ref{eqn:coherency_matrix}) leads to
bandwidth depolarization when the orientation of $\bar{\mbf{S}}$
varies with $\nu$.

An arbitrary $2\times2$ matrix may also be represented by the polar
decomposition,
\begin{equation}
{\bf J} = J \; \boost \rotat,
\label{eqn:polar_decomposition}
\end{equation}
where $J=(\det{\bf J})^{1/2}$,
\begin{eqnarray}
\boost &=& \exp(\boldsymbol{\sigma\cdot}\mbf{\hat m}\beta)
	= [\cosh\beta,\sinh\beta\;\mbf{\hat m}],
\label{eqn:boost} \\
\rotat &=& \exp(i\boldsymbol{\sigma\cdot}\mbf{\hat n}\phi)
	= [\cos\phi,i\sin\phi\;\mbf{\hat n}],
\label{eqn:rotation}
\end{eqnarray}
and $\mbf{\hat m}$ and $\mbf{\hat n}$ are real-valued unit 3-vectors.
The matrix, \rotat, is unitary and, beginning with
equation~(\ref{eqn:congruence_transformation}), it can be shown to
preserve the degree of polarization and rotate $\mbf{S}$ about the
axis, $\mbf{\hat n}$, by an angle $2\phi$.  Similarly, the Hermitian
matrix, \boost, can be shown to perform a Lorentz boost on the
4-vector, [$S_0$, $\mbf{S}$], along the axis, $\mbf{\hat m}$, by a
velocity parameter $2\beta$ \citep{bri00}.

As both \boost\ and \rotat\ are unimodular, the congruence
transformation (eq.~[\ref{eqn:congruence_transformation}])
preserves the determinant up to a multiplicative constant, $|J|^2$,
ie. $\det\bar{\bf P}^\prime=|J|^2\det\bar{\bf P}$.  Britton notes that
$\det\bar{\bf P}=\bar{S}_0^2-|\bar{\mbf{S}}|^2=\bar{S}^2_{\rm inv}$ is
simply the Lorentz invariant, which he calls the polarimetric
invariant interval.  He also proposes that a mean pulsar profile
formed using the invariant interval may be used for high-precision
pulsar timing.  Whereas the boost component of the system response
distorts the total intensity profile in a time-dependent manner, the
invariant interval remains stable, providing a superior basis for
arrival time estimates and a robust alternative to polarimetric
calibration.  However, the invariant interval is not preserved in the
presence of bandwidth depolarization, further motivating the current
development.

\section{Determination of the System Response}
\label{sec:system_response}

In principle, any method by which the frequency response matrix may be
derived with sufficiently high resolution could be adapted for use
with the matrix convolution approach.  A number of different
parameterizations and techniques have previously been presented for
the calibration of single-dish radio astronomical instrumental
polarization \citep{scr+84,tfsw85,xil91,bri00}.  The current treatment
follows the parameterization chosen by \citet{ham00}, including polar
decomposition of the system response followed by application of the
congruence transformation.  This approach offers a number of
significant advantages.  For instance, in the derivation of the
solution, it is unnecessary to make any small-angle approximations
and, when compared to the manipulations of 4$\times$4 Mueller
matrices, the equivalent quaternion and 2$\times$2 Jones matrix
representations greatly simplify the required algebra.  As well,
equations~(\ref{eqn:congruence_transformation})
and~(\ref{eqn:polar_decomposition}) are specific to no particular
basis, permitting the application of the formalism to a variety of
feed designs by proper choice of the basis matrices,
$\boldsymbol{\sigma}$.  Perhaps most importantly, as shown by Hamaker
and discussed below, the polar decomposition enables the determination
of the boost component of the system response using only an
observation of an unpolarized source.  This simplification has
beneficial impact on experiments where only the total intensity or
fractional degree of polarization are of concern.


Consider the reception of unpolarized radiation, which has an input
average power spectrum matrix, $\bar{\bf{P}}_L$=${\bf I}\;T_0/2$, where
${\bf I}$ is the identity matrix and $T_0$ is the total intensity.
Beginning with equations~(\ref{eqn:congruence_transformation})
and~(\ref{eqn:polar_decomposition}), the output average power spectrum
can be trivially shown to be
\begin{equation}
{\bar{\bf{P}}^\prime_L}=\boost^2 \;|J|^2 T_0/2.
\label{eqn:boost_soln}
\end{equation}
Notice that the phase of $J$ is lost in the detection of the average
power spectrum.  For this reason, the current technique is insensitive
to absolute phase terms arising in the frequency response of the
system, and may be used to determine only the relative phase
differences between its components.  The scalar factors, $|J|^2$ and
$T_0$, are easily determined in a separate flux calibration procedure
that will not be considered presently.  Therefore, \boost\ may be
found by taking the positive Hermitian square root of the measured
average power spectrum matrix of an unpolarized source 
(see Appendix~\ref{app:solve}).

This simple result merits closer inspection.  The system response may
be corrected by inverting equation~(\ref{eqn:convolution_nu}), that is,
by calculating $\mbf{E}={\bf J}^{-1}\mbf{E^\prime}$.  Notice that,
under the polar decomposition chosen in
equation~(\ref{eqn:polar_decomposition}), the boost transformation is
the first to be inverted.  That is, regardless of the unknown
rotation, the boost solution may be used to completely invert the
distortion of total intensity, $S_0$.  This consequence of the polar
decomposition has considerable impact in the field of high-precision
pulsar timing, where the average total intensity profile is used to
determine the pulse time of arrival.  Polarimetric distortions to the
total intensity can significantly alter the shape of this profile and
systematically alter arrival time estimates, especially as a function
of parallactic angle, or with fluctuations of ionospheric total
electron content.  With the boost component thus determined, the
resulting distortions to total intensity can, in principle, be
completely corrected.

It remains to solve for the rotation component of {\bf J}, which may
be determined using observations of calibrators with known
polarization.  Given the known boost, \boost, input polarization
state, $\bar{\bf P}_n$, and measured output state, $\bar{\bf
P}^\prime_n$, equation~(\ref{eqn:congruence_transformation}) is solved
for the rotation, \rotat.  Considering the equivalent
three-dimensional Euclidian rotation, \rotEn, of the polarization
vector, $\mbf{S}$, it is easily seen that, given a single pair of
input, $\mbf{S_1}$, and output, $\mbf{S_1}^\prime$, polarization
vectors, the rotations that solve $\mbf{S_1}^\prime=\rotEn\mbf{S_1}$,
where $|\mbf{S_1}^\prime|=|\mbf{S_1}|$, form an infinite set of
rotations with axis confined only to a plane.  Therefore, an
observation of a second, non-collinear calibrator source is required
in order to uniquely determine the system response rotation.

Many receivers are equipped with a linearly polarized noise diode that
may be used to inject a calibrator signal into the feed horn.  This
noise diode may be switched using a wide-band, amplitude-modulated
square-wave.  The ``off'' or ``low'' fraction of the wave consists of
only the system plus sky temperature, which shall be assumed to be
unpolarized.  The ``on'' or ``high'' fraction of the wave contains
additional linearly polarized radiation, described in quaternion form
by
\begin{equation}
\bar{\bf{P}}_H = [1,(\cos2\Psi,\sin2\Psi,0 )]\;C_0/2,
\label{eqn:high_cal}
\end{equation}
where $C_0$ is the flux and $\Psi$ is the position angle of the
calibrator diode.  A single linear noise diode provides only one known
input calibrator state.  Unless another calibrator is available, the
technique must be modified to solve for a reduced representation of
the system response.  For the calibration described in this paper,
\rotat\ was be decomposed into two rotations: \rotv, allowing
imperfect alignment of the noise diode; followed by \rotq, allowing a
differential path length between the two linear polarizations (see
Appendix~\ref{app:solve}).  It is important to distinguish
$\Delta\Psi$ and $\Phi_I$ from the actual parameters of the receiver,
as the polar decomposition does not model the order in which the 
transformations physically occur.

\subsection{Calibrator Observations}
\label{sec:system_response:cal}

When observing the artificial calibrator, it is important that the
noise diode be switched on a time-scale much shorter than the interval
over which the digitization thresholds are reset, in order that
$\bar{\bf{P}}_H$ may be differentiated from the baseline,
$\bar{\bf{P}}_L$.  However, the calibrator period must also be long
enough to provide distinct on-pulse and off-pulse time samples in the
synthesized filterbank.  Given the sampling interval, $t_s$, of the
baseband recorder, the calibrator period, $T_C$, should be at least
$T_C \ge n_b t_s N$, where $N$ is the number of channels in the
synthetic filterbank and $n_b$ is the desired number of phase bins in
the integrated calibrator profile.  In order to achieve clear
separation of on-pulse and off-pulse states, $n_b=64$ was chosen.  As
well, it was found that a filterbank with $N=2048$ channels
sufficiently resolved the features of the frequency response.

For each hour-long observation of \psr, the pulsed calibrator was
recorded for 4.5 minutes, ensuring sufficient signal-to-noise in each
channel of the synthetic filterbank.  Data were reduced offline using
{\tt psrdisp}, a software package developed to process pulsar baseband
data, described in the following section.  After forming a synthetic
filterbank, the Stokes parameters in each channel were detected and
folded at the pulsed calibrator period.  High and low state
polarimetric passbands were formed from the average on- and off-pulse
Stokes parameters of the calibrator profile in each channel.  These
passbands were median filtered to remove spurious radio-frequency
interference, and interpolated to the required frequency resolution,
as dictated by the parameters of the coherent dedispersion kernel.
From this high frequency-resolution polarimetric representation of the
calibrator, the frequency response matrix was computed as described in
Appendix~\ref{app:solve}.  Figure~\ref{fig:pcal} plots representative
examples of the determined frequency response parameters at 660\,MHz
and 1247\,MHz.  The inverse of the frequency response matrix was used
in the convolution kernel when reducing the pulsar observations, as
described in the next section.

\begin{figure*}[h]
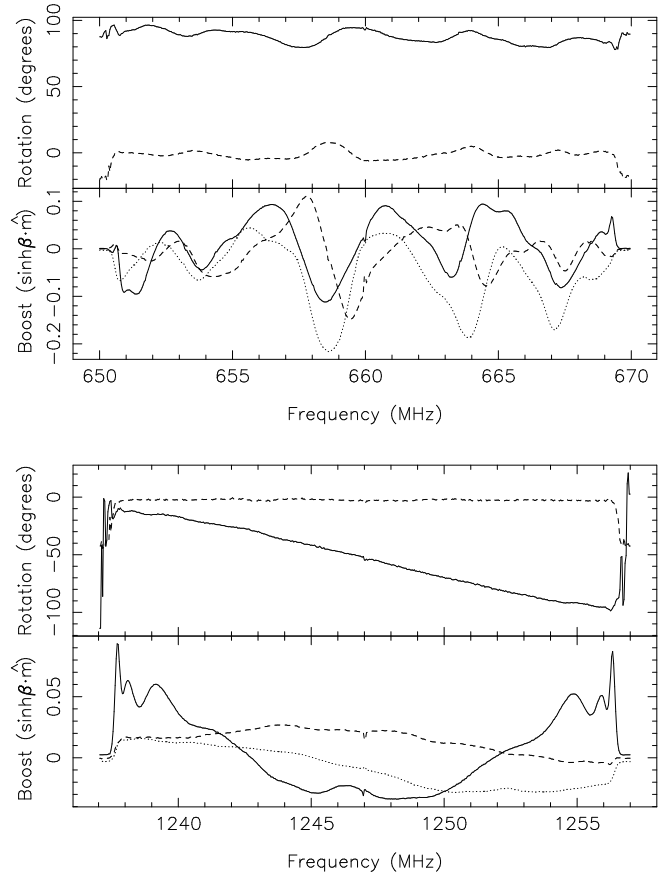

\centerline{\psfig{figure=pcal_660.ps,width=86mm,angle=-90}}
\vspace{5mm}
\centerline{\psfig{figure=pcal_1247.ps,width=86mm,angle=-90}}
\caption {Frequency response at 50\,cm (top) and 20\,cm (bottom).
For each band, the upper panel plots the rotation angles, $\Delta\Psi$
(dashed line) and $\Phi_I$ (solid line), and the lower panel plots the
boost components along $\mbf{\hat q}$ (solid line), $\mbf{\hat u}$
(dotted line), and $\mbf{\hat v}$ (dashed line).  A signal path length
mismatch between the two polarizations from the Multibeam (20\,cm)
receiver results in a linear differential phase ($\Phi_I$) gradient
across the bandpass.}
\label{fig:pcal}
\end{figure*}

\section{Pulsar Observations}

\begin{figure*}[h]
\centerline{\psfig{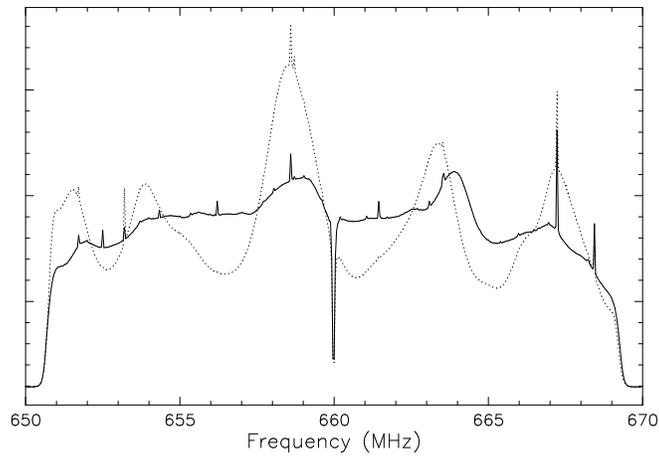}}
\caption {Average passbands from the two linear polarizations at
660\,MHz.  The gain in $E_y$ (dotted line) varies wildly with
frequency, evidencing the less than ideal condition of the 50\,cm
receiver.}
\label{fig:pband_660}
\end{figure*}

Dual-polarization baseband data from the nearby millisecond pulsar,
\psr, were recorded during 2001 June 26-28, using the 64\,m dish at
the Parkes Observatory.  As a test of phase-coherent calibration in
the pathological case (refer to Figure~\ref{fig:pband_660}), the
50\,cm receiver was used to record 3.6 hours of data at 660\,MHz.  As
well, a total of 7 hours was recorded at 1247\,MHz, using the centre
element of the Multibeam receiver.  The radio and intermediate
frequency signals were quadrature downconverted, band-limited to 20
MHz, two-bit sampled, and recorded using the Caltech-Parkes-Swinburne
Recorder (CPSR) \citep{vbb00}.  The digitized data were processed
offline at Swinburne University's supercomputing facilities using {\tt
psrdisp}, a flexible software package that implements a number of
baseband data reduction options.

The four channels of two-bit quantized data were corrected using the
dynamic level-setting technique \citep{ja98} and combined to form the
digitized analytic vector, $\mbf{e}(t_i)$.  This vector was convolved
while synthesizing a 16-channel coherent filterbank \citep{jcpu97}, a
process described in more detail in Appendix~\ref{app:convolve}.  In
each channel, the Stokes 4-vector, $[S_0,\mbf{S}](\nu_k, t_n)$, was
detected and integrated as a function of pulse phase, given by $t_n$
modulo the predicted topocentric period.  Pulse period and absolute
phase were calculated using a polynomial generated by the {\tt tempo}
software package (http://pulsar.princeton.edu/tempo).

\begin{figure*}
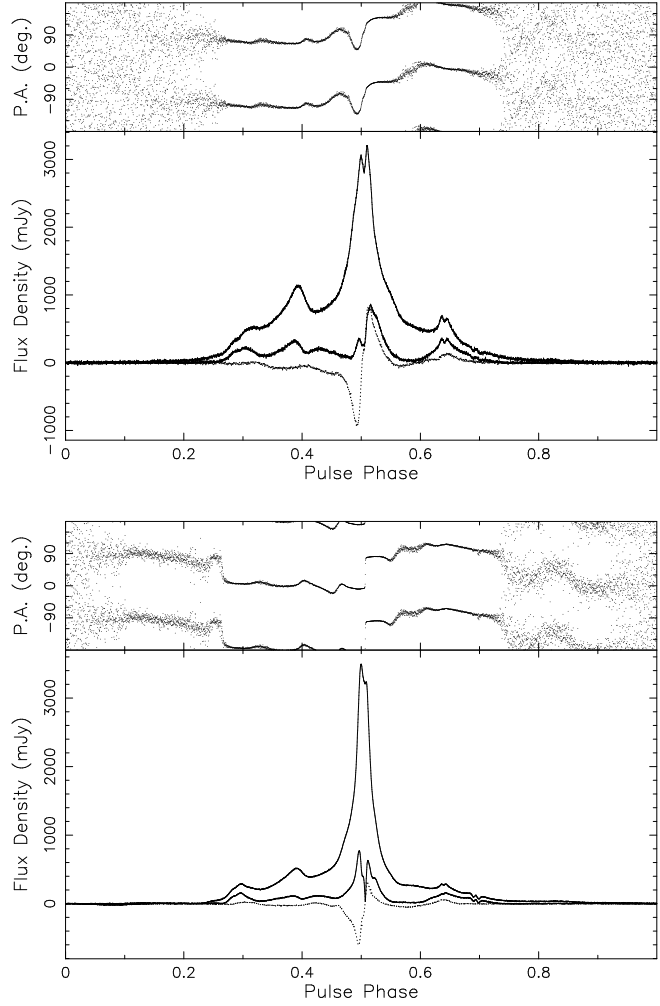

\centerline{\psfig{figure=0437_660.ps,width=86mm,angle=-90}}
\vspace{5mm}
\centerline{\psfig{figure=0437_1247.ps,width=86mm,angle=-90}}
\caption {Average polarimetric pulse profiles of \psr\ at 660\,MHz
(top) and 1247\,MHz (bottom).  The bottom panel of each plot displays
the total intensity, Stokes $I$ (upper solid line), the linearly
polarized component, $L$ (lower solid line), and the circularly
polarized component, Stokes $V$ (dotted line).  The position angle,
$\psi$, in the upper panel of each plot is shown without respect
to any absolute frame of reference.}
\label{fig:0437}
\end{figure*}

\begin{figure*}
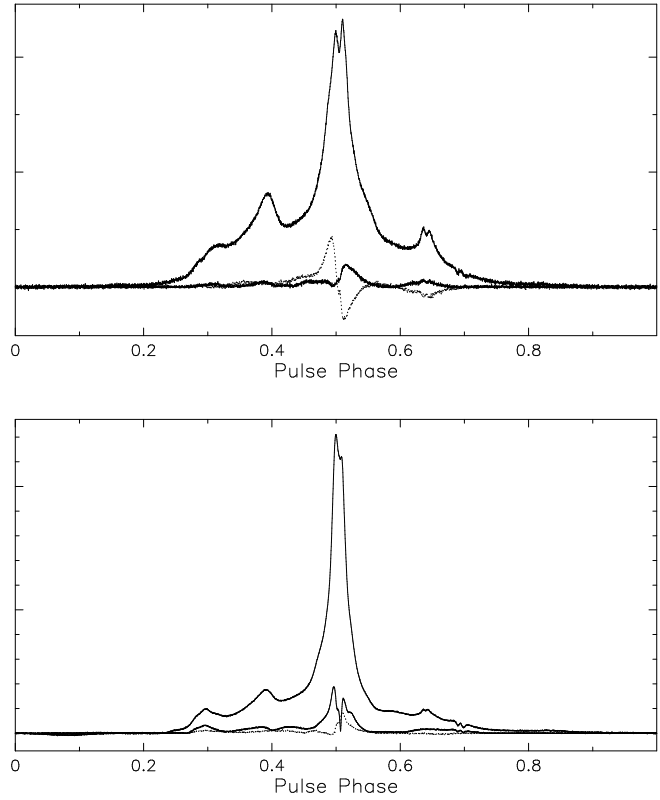

\centerline{\psfig{figure=uncal_660.ps,width=86mm,angle=-90}}
\vspace{5mm}
\centerline{\psfig{figure=uncal_1247.ps,width=86mm,angle=-90}}
\caption {Uncalibrated average polarimetric pulse profiles of \psr\
at 660\,MHz (top) and 1247\,MHz (bottom).  Severe differential gain
variation in the 50\,cm bandpass (see Figure~\ref{fig:pband_660}) has
decimated the linearly polarized component at 660\,MHz (top plot,
lower solid line), whereas the differential phase gradient across the
20\,cm bandpass (see Figure~\ref{fig:pcal}) has nearly eliminated the
circularly polarized component at 1247\,MHz (bottom plot, dotted
line).}
\label{fig:uncal}
\end{figure*}

Each 1\,GB segment of baseband data (representing approximately 53.7
seconds) was processed in this manner, producing average pulse profiles
with 4096 phase bins, or a resultant time resolution of approximately
1.4\,$\mu$s.  Each archive was later corrected for parallactic angle
rotation of the receiver feeds before further integrating in time to
produce hour-long average profiles.  These were flux calibrated using
observations of Hydra\,A, which was assumed to have a flux density of
approximately 85\,Jy at 660\,MHz and 48\,Jy at 1247\,MHz.  The flux
calibrated archives were then integrated in time and frequency,
producing the mean pulse profiles presented in Figure~\ref{fig:0437}.
In these plots, the linearly polarized component, $L$, is given by
$L^2=Q^2+U^2$ and the position angle, $\psi$, where $\tan\psi=U/Q$, has been
plotted twice, at $\psi\pm\pi/2$.  Comparison with the uncalibrated
profiles in Figure~\ref{fig:uncal} indicates significant restoration
of the polarized component at both frequencies.  As well, the
calibrated average profiles agree satisfactorily with previously
published polarization data for \psr\ \citep{nms+97}.

\section {Discussion}

Prior to its reception, the radiation from a pulsar must propagate
through the magnetized plasma of the ISM, resulting in Faraday
rotation of the position angle, $\psi$.  The change in position angle
is given by $\Delta\psi=RM\lambda^2$, where $RM$ is the rotation
measure and $\lambda$ is the wavelength.  Therefore, one motivation
for a pulsar polarimetry experiment is the determination of the RM
along the line of sight to the pulsar.  Combined with a dispersion
measure estimate, the RM can provide information about the galactic
magnetic field strength and direction.  When the RM is very small, as
it is for \psr, $\Delta\psi$ is not detectable across a single 20\,MHz
bandpass.  Therefore, two observations widely separated in frequency
are required in order to obtain an estimate.  However, as there exists
an unknown feed offset between 50\,cm and Multibeam receivers that
remains uncalibrated in the current treatment, no estimation of
rotation measure is presently offered.

For pulsars with known RM, matrix convolution enables phase-coherent
Faraday rotation correction by addition of $\Delta\psi$ to
$\Delta\Psi$ in equation~(\ref{eqn:response}).  This approach would
further treat the problem of bandwidth depolarization, especially for
sources with large rotation measures observed at lower frequencies.

As phase-coherent calibration corrects the undetected voltages, any
measurements derived from these data will also be implicitly
calibrated.  This includes not only the coherency products but also
other statistical values such as the higher order moments, modulation
index, and polarization covariance.  The method may therefore find
application in experiments which aim to describe pulse shape
fluctuations \citep{jap01} or variations of the galactic magnetic
field along the line of sight \citep{mm98}, for example.

Although matrix convolution may in principle be used to completely
correct baseband data, a full solution of the system response is
lacking in the current treatment.  Perhaps a technique may be
developed which combines the use of pulsars \citep{scr+84,xil91}
with high-resolution determination of the frequency response matrix.
This approach would include tracking a bright, strongly polarized
pulsar over a wide range of parallactic angles, followed by fitting a
model of the instrumental response to the measured Stokes parameters.
As it is non-trivial to model the effects of phenomena that may change
over the course of such a calibration, such as fluctuations in
ionospheric total electron content, it may prove more feasible to
install a second artificial noise diode in the receiver feed horn.
Oriented with a position angle offset from the first by approximately
45\degr, it would provide a more readily available and reliable means
of regularly solving for \rotat.  Alternatively, a single noise diode
could be mechanically rotated with respect to the feeds, obviating the
inevitable cross-talk that would arise between two diodes.
Consideration of these options may benefit future feed horn design.

The observations of \psr\ presented in the previous section serve to
illustrate the strength the matrix convolution technique, highlighting
its ability to recover data which has been severely corrupted by the
observatory instrumentation.  Perhaps the most rigorous test of the
methodology will be derived from the unsurpassed timing accuracy of
\psr\ \citep{vbb+01}.  Calibration errors translate into systematic
timing errors, and any deficiency in the characterization of the
system response would manifest itself in the arrival time residuals of
this remarkable pulsar.

\vspace{-4mm}
\acknowledgements
\small
The Parkes Observatory is part of the Australia Telescope which is
funded by the Commonwealth of Australia for operation as a National
Facility managed by CSIRO.  This research was supported by the
Commonwealth Scholarship and Fellowship Plan.  I am grateful to
Matthew Britton and Simon Johnston, with whom I have shared many
stimulating discussions on radio polarimetry.  Thanks also to Matthew
Bailes and Stephen Ord for assistance with observing and helpful
comments on the text.
\normalsize

\appendix

\section{Solution of System Response}
\label{app:solve}

The instrumental frequency response matrix used in calibrating the
data presented in this paper was parameterized by
\begin{equation}
{\bf J} = \boost\rotq\rotv.
\label{eqn:response}
\end{equation}
The boost component may be solved most easily using the
quaternion form of equation~(\ref{eqn:boost_soln}),
\begin{equation}
[\bar{L}_0^\prime,\bar{\mbf{L}}^\prime]
	= \exp(2\boldsymbol{\sigma\cdot}\mbf{\hat m}\beta) |J|^2 T_0
	= [\cosh(2\beta),\sinh(2\beta)\mbf{\hat m}]\;|J|^2 T_0,
\label{eqn:boost_quaternion}
\end{equation}
where $[\bar{L}_0^\prime,\bar{\mbf{L}}^\prime]$ are the measured off-pulse
Stokes parameters, producing
\begin{equation}
\beta=\frac{1}{2}\tanh^{-1}\left({{|\bar{\mbf{L}}^\prime|}
			\over      {\bar{L}_0^\prime}
				  }\right)
\hspace{1cm} {\rm and} \hspace{1cm} 
\mbf{\hat m}= {{\bar{\mbf{L}}^\prime}\over{|\bar{\mbf{L}}^\prime|}}.
\label{eqn:boost_solve}
\end{equation}
The two rotations, \rotq\ and \rotv, may be determined by considering
the equivalent three-dimensional Euclidian rotations, \rotEq\ and
\rotEv, of the input polarization vector.  Given that the noise diodes
installed in the receivers at the Parkes radio-telescope have a
position angle nearly equal to 45\degr, the input Stokes parameters
are $[1,\bar{\mbf{H}}]C_0$, where $\bar{\mbf{H}}=(0,1,0)$
(cf. eq.~[\ref{eqn:high_cal}]).  Therefore, $\Phi_I$ and $\Delta\Psi$
may be found by solving
\begin{equation}
\bar{\mbf{H}}^{\prime\prime}=\rotEq\rotEv\bar{\mbf{H}},
\label{eqn:rotate_euclid}
\end{equation}
where $\bar{\mbf{H}}^{\prime\prime}$ is the normalized polarization
vector after the observed Stokes parameters have been corrected for 
the boost.  It is given by
\begin{equation}
[\bar{H}_0,\bar{\mbf{H}}]^{\prime\prime}
=\boost^\dagger\bar{\bf{P}}_H^{\prime}\boost/(|J|^2C_0),
\end{equation}
where $\bar{\bf{P}}_H^{\prime}$ are the observed Stokes parameters.
From equation~(\ref{eqn:rotate_euclid}),
\begin{equation}
\Phi_I=\frac{1}{2}\tan^{-1}\left({{-H_3^{\prime\prime}}
			\over      {H_2^{\prime\prime}}
				  }\right)
\hspace{1cm} {\rm and} \hspace{1cm} 
\Delta\Psi=\frac{1}{2}\tan^{-1}\left({{H_1^{\prime\prime}}
			\over          {H_2^{\prime\prime}/\cos(2\Phi_I)}
				      }\right).
\label{eqn:rotate_solve}
\end{equation}
Notice that $|J|^2$, $T_0$, and $C_0$ cancel out in each of
equations~(\ref{eqn:boost_solve}) and~(\ref{eqn:rotate_solve}),
obviating the need to solve for these parameters at this stage.  The
vector components of the boost quaternion,
$(B_1,B_2,B_3)=\sinh\beta\,\mbf{\hat m}$, as well as $\Phi_I$ and
$\Delta\Psi$, are plotted as a function of frequency in
Figure~\ref{fig:pcal}.

\section{Performing Matrix Convolution}
\label{app:convolve}

It is assumed that the reader has some familiarity with the more
common, one-dimensional form of cyclical convolution as it is
performed in the frequency domain using the Fast Fourier Transform
(FFT) \citep[ \S 13.1]{ptvf92}.  In the two-dimensional vector
case, there are simply two unique processes sampled at the same time
interval, say $p(t_i)$ and $q(t_i)$.  A one-dimensional, $N$-point,
forward FFT is performed separately on each of $p$ and $q$, forming
two spectra, $P(\nu_k)$ and $Q(\nu_k)$, $0\le k<N$.  Corresponding
elements from each of the spectra are treated as the components of a
column 2-vector, $\mbf{E}(\nu_k)=(P(\nu_k),Q(\nu_k))$, and multiplied
by the inverse of the frequency response matrix, forming
$\mbf{E}^\prime(\nu_k)={\bf{J}}^{-1}(\nu_k)\mbf{E}(\nu_k)$
(cf. eq.~[\ref{eqn:convolution_nu}]).  The components of the result,
$P^\prime(\nu_k)$ and $Q^\prime(\nu_k)$, are once again treated as
unique spectra, and separately transformed back into the time domain
using the one-dimensional backward FFT.

In the case of the pulsar data presented in this paper, $p(t_i)$ and
$q(t_i)$ are the signals from the two linear feeds in the receiver,
and ${\bf{J}}(\nu_k)$ consists of the instrumental frequency response
matrix, as determined in Appendix~\ref{app:solve}, multiplied by the
dispersion kernel, $H(\nu_k)$ \citep{hr75}.  When synthesizing an
$M$-channel coherent filterbank, $H(\nu_k)$ is divided in frequency
into $M$ distinct dispersion kernels, each tuned to the centre
frequency of the resulting filterbank channel.


\begin{thebibliography}{}

\bibitem[Britton(2000)]{bri00}
Britton, M.~C. 2000, ApJ, {532}, 1240

\bibitem[{Gangadhara} {\it et~al.}(1999)]{gxv+99}
{Gangadhara}, R.~T., {Xilouris}, K.~M., {von Hoensbroech}, A., {Kramer}, M.,
  {Jessner}, A., \& {Wielebinski}, R. 1999, A\&A, 342, 474

\bibitem[Hamaker(2000)]{ham00}
Hamaker, J.~P. 2000, A\&AS, 143, 515

\bibitem[{Hankins} \& {Rickett}(1975)]{hr75}
{Hankins}, T.~H., \& {Rickett}, B.~J. 1975, in Methods in Computational
  Physics 14, Radio Astronomy (New York: Academic Press), 55

\bibitem[Jenet \& Anderson(1998)]{ja98}
Jenet, F.~A., \& Anderson, S.~B. 1998, PASP, 110, 1467

\bibitem[{Jenet}, {Anderson}, \& {Prince}(2001)]{jap01}
{Jenet}, F.~A., {Anderson}, S.~B., \& {Prince}, T.~A. 2001, ApJ, 546, 394

\bibitem[Jenet {\it et~al.}(1997)]{jcpu97}
Jenet, F.~A., Cook, W.~R., Prince, T.~A., \& Unwin, S.~C. 1997, PASP, 109, 707

\bibitem[Melrose \& Macquart(1998)]{mm98}
Melrose, D., \& Macquart, J. 1998, ApJ, 505, 921

\bibitem[Navarro {\it et~al.}(1997)]{nms+97}
Navarro, J., Manchester, R.~N., Sandhu, J.~S., Kulkarni, S.~R., \& Bailes, M.
  1997, ApJ, 486, 1019

\bibitem[Papoulis(1965)]{pap65}
Papoulis, A. 1965, {Probability, Random Variables, and Stochastic
  Processes} (Sydney: McGraw-Hill)

\bibitem[Press {\it et~al.}(1992)]{ptvf92}
Press, W.~H., Teukolsky, S.~A., Vetterling, W.~T., \& Flannery, B.~P. 1992,
  {Numerical Recipes: The Art of Scientific Computing}
  (2$^{nd}$\,ed.; Cambridge: Cambridge University Press)

\bibitem[Stinebring {\it et~al.}(1984)]{scr+84}
Stinebring, D.~R., Cordes, J.~M., Rankin, J.~M., Weisberg, J.~M., \&
  Boriakoff, V. 1984, ApJS, 55, 247

\bibitem[Turlo {\it et~al.}(1985)]{tfsw85}
Turlo, Z., Forkert, T., Sieber, W., \& Wilson, W. 1985, A\&A, 142, 181

\bibitem[van Straten, Britton, \& Bailes(2000)]{vbb00}
van Straten, W., Britton, M., \& Bailes, M. 2000, in ASP Conf. Ser. 202,
{Pulsar Astronomy - 2000 and Beyond}, ed.\ M. Kramer, N. Wex, \& R.
Wielebinski (San Francisco: ASP), 283

\bibitem[van Straten {\it et~al.}(2001)]{vbb+01}
van Straten, W., Bailes, M., Britton, M., Kulkarni, S.~R.,
Anderson, S.~B., Manchester, R.~N., \& Sarkissian, J. 2001, Nature, 412, 158,
astro-ph/0108254

\bibitem[Xilouris(1991)]{xil91}
Xilouris, K.~M. 1991, A\&A, 248, 323

\end{thebibliography}
\end{document}